\documentclass[aps,pra,nofootinbib,preprint,amsmath,amssymb,floatfix]{revtex4}
\usepackage{color}
\usepackage{graphicx}

\begin{document}

\newcommand{\g}{blue}
\newcommand{\ve}{\varepsilon}
\title{Uniform quantized electron gas: Radiation corrections}         

\author{Johan S. H{\o}ye$^1$ and Enrique Lomba$^2$}      
\affiliation{$^1$Department of Physics, Norwegian University of Science and Technology, N-7491 Trondheim, Norway\\
$^2$ Instituto de Q{\'\i}mica F{\'\i}sica Rocasolano, CSIC, Serrano 119, E-28006 Madrid, Spain}
\date{\today}          

\begin{abstract}
In this paper we analyze how radiation effects influence the
correlation functions, the excess energy, and in turn the electron correlation energy 
of the quantized electron gas at temperature $T=0$. 
To that aim we resort to a statistical mechanical
description of the quantum problem of electron correlations, based on
the path integral formalism. In previous works we studied and found accurate results
for the usual situation with the electrostatic Coulomb interaction.
Here the additional problem with radiation is taken into account. 
This is facilitated by the equivalence to a dielectric fluid for which correlation functions 
for dipolar moments are established. From these functions follows the usual density-density
(or charge-charge) correlation function needed for the longitudinal electrostatic problem,
and in addition the one needed for the transverse radiation problem. While electrostatic 
excess energy is negative, the transverse one is positive. This quantity is small and decreases rapidly 
for decreasing densities. However, for high densities it approaches
the electrostatic contribution, eventually becoming even larger. The
part of the transverse energy from induced correlations  
turns out to be very small. Also, the non-local longitudinal and transverse dielectric constants of the electron gas are 
identified from the induced correlation functions.

  \end{abstract}
\maketitle

\bigskip
\section{Introduction}
\label{secintro}
Electron correlation energy is a fundamental quantity in quantum
chemistry calculations. Standard procedures for its evaluation are
Configuration Interaction approaches or many-body perturbation
theories \cite{Fulde1993}. In a series of works
\cite{hoye10,hoye16,lomba17} the authors have developed an alternative
approach based on the standard procedures of classical statistical mechanics
using the path integral formalism, which 
translates the quantum problem into a classical polymer problem in four dimensions. 
The well known RPA (random phase approximation) was recovered as a
basic result \cite{hoye16}.  Then in Ref.~\cite{lomba17} a much more accurate
approximation was derived starting from  the enforcement of thermodynamic
consistency.

In this paper we analyze and evaluate how radiation effects influence  
the excess and correlation energies of the uniform quantized electron gas.
These results can also provide an indication of the
significance of these effects  for electrons in molecules.

In this connection,  quantum computations require the use of non-zero frequencies of the interaction.
This implies that  the Coulomb interaction has to be extended in a
non-trivial way, since  it must involve the electromagnetic
vector potential. This again means that current-current correlations have
to be taken into account. This situation was analyzed in Ref.~\cite{hoye10}
where it was found that the problem can be simplified by introducing polarizations,
through which charge densities and currents can be expressed. In this way the problem
is transformed into that of a dielectric fluid with radiating dipole-dipole interaction.

As shown in Sec.~\ref{sec3} the dipole interaction can be separated into two components
that do not couple. This simplifies considerably in our case since one
part will recover precisely the standard electrostatic result, by which
the remaining contribution is entirely due to radiation. This separation of the dipolar interaction has been
utilized earlier in theories for classical dielectric fluids. It was already used by 
Wertheim to solve the MSA (Mean Spherical Approximation) for dipolar hard
spheres where both terms are present in the electrostatic
case~\cite{wertheim71,hoye74} as well. The same two terms also occur
in the study of the refractive index of polarizable fluids with radiating interaction 
\cite{hoye82}. Note that this  uncoupling is just a particular case of the general
rotational invariant expansion proposed by Blum and
Torruella~\cite{blum72} to deal with two-body correlations in
molecular fluids. 

As already mentioned, one part of the interaction and the resulting correlation function of the electron gas
is tied to the longitudinal electrostatic interaction. The other part
is connected to the transverse electromagnetic radiation. The first
part, the usual electrostatic problem, was studied in
Refs.~\cite{hoye16,lomba17} where it was shown that the approach leads
accurate results. In the present case the additional contribution from
 the transverse frequency dependent interaction is included. This is
 obtained by extension of the statistical mechanical method introduced
 in Refs.~\cite{hoye16,lomba17}. This method also coincides with the
 well-known RPA  when effective
 interaction is not used.    
The additional problem with the transverse contribution is the need for current correlations for the free electron gas. However, expressions for these follow
from Ref.~\cite{hoye10}.

By our computations we find that radiations effects are relatively small
compared to the the contribution due to the usual electrostatic interaction
alone. This is consistent with the previous study of polarizable hard spheres where 
the positions of the spheres were treated classically while their oscillating
dipole moments were quantized \cite{waage13}. This is also consistent with
previous results for induced Casimir energies that are known to be
responsible for the  attraction between dielectric media. 
These energies are usually attributed to quantum fluctuations of the electromagnetic field \cite{casimir48}.
But this field can be eliminated to be replaced with radiating dipole interaction between
polarizable particles \cite{brevik88}. For a pair of polarizable
particles one, for the Casimir energy, finds an 
induced $1/r^6$ van der Waals interaction for small separations while for large separations the faster $1/r^7$ decay takes place due to retardation (radiation) effects \cite{casimir48,brevik88}. Thus, radiation reduces the induced interaction
for large $r$, but when integrating over it
the total net effect of radiation is small.
Anyway, our analysis of the 
transverse exchange energy shows that the influence of
radiation effects is most visible for high electron densities.

The rest of the paper is organized as follows. In Section \ref{sec2}
we briefly review the expressions that define the correlation
functions for the electron gas in the path integral formalism. The
radiating dipole interaction potential is presented in Section
\ref{sec3}. In Sections \ref{exc} and \ref{corr} we find the
exchange and correlation electron energies. First we recover previous expressions for the longitudinal electrostatic case. Then these expressions are extended to the new situation where radiation is included by which the additional transverse contribution is obtained. The 
dielectric constant of the electron gas is discussed in Section \ref{sec8}. Finally in
Section \ref{sec9} numerical results are presented. The expressions
for analytic integration of the transverse exchange energy are
explicitly detailed in the Appendix.

\section{Correlation functions}
\label{sec2}

The charge-charge (or density-density) correlation function of the free fermion gas and the electrostatic Coulomb potential were needed to obtain the correlation energy of the uniform electron gas. However, there will be radiation corrections that are expected to be small. These we will incorporate in a quantitative way by extension of the method used to obtain our previous results for the correlation energy. This extension will require current-current correlations to take into account the electromagnetic vector potential that appears when time-dependent electromagnetic forces are taken into account. This has been studied on a more general basis in Ref.~\cite{hoye10}. There it was noticed that the electron problem  can be regarded and transformed to the problem of a dielectric fluid. 

The transformation to a dielectric fluid simplifies the problem since both charge density $\rho$ and current density ${\bf j}$ can be expressed  in terms of polarization ${\bf P}$. One has 
\begin{equation}
\rho=-\nabla {\bf P}\quad \mbox{and}\quad {\bf j}=\frac{\partial {\bf P}}{\partial t}.
\label{10}
\end{equation}
This is Eq.~(I18) of Ref.~\cite{hoye10}. For simplicity equations of this reference are preceded by the numeral I when referred to. With this Eq.~(I14), the equation of charge conservation $\partial \rho/\partial t+\nabla{\bf j}=0$ is fulfilled. By Fourier transform  (signs depending upon convention used)
\begin{equation}
\nabla\rightarrow-i{\bf k}, \quad \frac{\partial}{\partial t}\rightarrow i\omega.
\label{11}
\end{equation}

The Fourier transform of the correlation function in imaginary time and space for a pair of polarizations ${\bf s}_1$ and ${\bf s}_2$ of the uniform free electron gas is given by Eq.~(I67) (for unit charges, $i,j=x,y,z$)
\begin{equation}
\langle s_{1i} s_{2j}\rangle=\hat P_{ij}(K,{\bf k})=\int\left[\frac{\hbar}{2m}(k_i^{''}-k_i^{'})\right]\left[\frac{\hbar}{2m}(k_j^{''}-k_j^{'})\right]\frac{\hbar^2}{\Delta^2+K^2}L\,d{\bf k}^{'},
\label{13}
\end{equation} 
\begin{equation}
L=\frac{\zeta}{(2\pi)^3}\frac{X-Y}{\Delta(1\pm \zeta X)(1\pm\zeta Y)}.
\label{14}
\end{equation}
The Matsubara frequency $K=i\hbar\omega$ ($\omega$ is frequency) is the Fourier variable in imaginary time. Further $\zeta=\exp(\beta\mu)$ where $\mu$ is chemical potential, $\beta=1/(k_B T)$, and  $k_B$ is Boltzmann's constant. In  the expression for $L$ the $+$ sign is for fermions. Further
\begin{equation}
X=F_\beta(k^{'}), \quad Y=F_\beta(k^{''}),\quad\mbox{with}\quad F_\lambda(k)=\exp{(-\lambda E(k))}, 
\label{15}
\end{equation}
\begin{equation}
E(k)=\frac{(\hbar k)^2}{2m}, \quad \Delta =E(k^{''})-E(k^{'}),\quad {\bf k}^{''}={\bf k}-{\bf k}^{'}.
\label{16}
\end{equation}
The imaginary time $\lambda$ is limited to $0\leq\lambda\leq\beta$.
Details of the derivation of expression (\ref{13}) can be found in
Ref.~\cite{hoye10}. However, one notes that the known density-density
correlation function (I69) follows as (where $\sum_{ij}$ is implicit)
\begin{equation}
\hat S(K,k)=k_i k_j \hat P_{ij}(K,k)=\int\frac{\Delta^2}{\Delta^2+K^2}L\,d{\bf k}{'}.
\label{17}
\end{equation}
This is also Eq.~(5) in Ref.~\cite{lomba17} --Eq.~(II5)--. Hereafter
the numeral II will be used to designate the equations of
Ref.~\cite{lomba17}. Note that due to  symmetry one can perform the
substitution 
$1/(iK+\Delta)\rightarrow
[1/(\Delta+iK)+1/(\Delta-iK)]/2=\Delta/(\Delta^2+K^2)$. 

\section{Radiating dipole interaction}
\label{sec3}

The radiating dipole-dipole pair interaction between two dipole moments ${\bf s}_1$ and ${\bf s}_2$ (per unit charge) is given by Eq.~(I24)
\begin{eqnarray}
\nonumber
\hat\phi(12,\omega)&=&\hat\psi(k,\omega)\left[({\bf k}\cdot{\bf s}_1)({\bf k}\cdot{\bf s}_2)-\left(\frac{\omega}{c}\right)^2{\bf s}_1\cdot{\bf s}_2\right],\\
\hat\psi(k,\omega)&=&\frac{e^2}{\varepsilon_0(k^2-(\omega/c)^2)}.
\label{12}
\end{eqnarray}
Here SI units are used instead of Gaussian ones ($4\pi\rightarrow1/\varepsilon_0$).
The $-e$ is the electric charge of the electron, $\varepsilon_0$ is the permittivity of vacuum, and $\omega$ is the frequency.

Now one notes that interaction (\ref{12}) can be written in the form
\begin{equation}
\hat\phi(12,\omega)=\tilde\psi(k)\left[J_1-\frac{(\omega/c)^2}{k^2-(\omega/c)^2}J_2\right],\quad \mbox{with}\quad \tilde\psi(k)=\frac{e^2}{\varepsilon_0 k^2},
\label{18}
\end{equation}
\begin{equation}
J_1=({\bf k}\cdot{\bf s}_1)({\bf k}\cdot{\bf s}_2), \quad J_2=k^2 {\bf s}_1\cdot{\bf s}_2-({\bf k}\cdot{\bf s}_1)({\bf k}\cdot{\bf s}_2)=({\bf k}\times{\bf s}_1)({\bf k}\times{\bf s}_2).
\label{19}
\end{equation}

With $\tilde\psi(k)$ alone we are back to the electrostatic Coulomb
interaction. To obtain the resulting correlation function and
correlation energy one has to perform convolutions as in the
electrostatic case of Ref.~\cite{lomba17}. As will be seen the
correlation function (\ref{13}) can be separated in $J_1$ and $J_2$
terms. A convolution (an orientational average) needed leads to
\begin{equation}
J_1 J_1=({\bf k}\cdot{\bf s}_1)\langle({\bf k}\cdot{\bf s}_3)({\bf k}\cdot{\bf s}_4)\rangle({\bf k}\cdot{\bf s}_2)=S_L J_1, 
\label{20}
\end{equation}
\begin{equation}
\langle J_1 \rangle=S_L, \quad S_L=\hat S_L(K,k)=\hat S(K,k),
\label{21}
\end{equation}
where  $\tilde S(K,k)$ is the correlation function (\ref{17}). Altogether with the $J_1$ term alone the RPA  of the electrostatic case is recovered. Also the $\tilde\psi(k)$ can be replaced by an effective (or cut) interaction as determined in Ref.~\cite{lomba17} to obtain accurate results.

To  include radiation the convolutions $J_1J_2$ and $J_2J_2$ are needed. One notes that from expression (\ref{13}) the matrix formed by $\hat P_{ij}=\hat P_{ij}(K,k)$ can be made diagonal. With the $z$ axis along the vector {${\bf k}=\{0,0,k\}$ one has $\tilde P_{ij}=0$ with $i\neq j$ and $\tilde P_{xx}=\tilde P_{yy}$ (since $k_x=k_y=0$). With this
\begin{equation}
J_1=k^2 s_{1z} s_{2z} \quad \mbox{and}\quad J_2=k^2(s_{1x} s_{2x}+s_{1y} s_{2y}).
\label{24}
\end{equation}
Accordingly $J_1 J_2=0$ and
\begin{equation}
J_2 J_2=k^4[s_{1x} \langle s_{3x}s_{4x}\rangle s_{2x}+s_{1y} \langle s_{3y}s_{4y}\rangle s_{2y}]=S_T J_2,
\label{26}
\end{equation}
\begin{equation}
\langle J_2\rangle=k^2[\langle s_{1x} s_{2x}\rangle+\langle s_{1y} s_{2y}\rangle]=2S_T, \quad S_T=\hat S_T(K,k)=k^2\hat P_{xx}=k^2\hat P_{yy}.
\label{27}
\end{equation}

Since $J_1 J_2=0$ the contribution from radiation decouples from the electrostatic result. By that the computations can be performed in a way similar to the electrostatic part, but with an interaction that follows from Eq.~(\ref{18}) and a reference system correlation function given by $\hat S_T(K,k)$ which couples to the transverse radiating field. Likewise $S_L=\hat S_L(K,k)$ is connected to the longitudinal electrostatic field as before.

\section{Exchange energies}
\label{exc}
In this section we will provide expressions to evaluate the exchange
energy contribution to the electron energy, due to both electrostatic and
radiation contributions. 
\subsection{Electrostatic exchange energy}
\label{sec4}

The exchange energy follows by integrating the interaction together with the correlation function and divide by two to avoid double counting. For the electrostatic case the equal (imaginary) time correlation function $\tilde S_L(0,k)=\tilde S(0,k)$ is needed. Also the self correlation given by particle density $\rho$ should be subtracted to avoid divergence. The electrostatic exchange energy per unit volume is thus
\begin{equation}
F_{Lex}=\frac{1}{2(2\pi)^3}   \int (g\tilde S(0,k)-\rho) \tilde\psi(k)\,d{\bf k}
\label{52}
\end{equation}
where $g=2$ is the spin degeneracy. With $g\tilde S(0,k)=\rho(3Q/2-Q^3/2)$ for $Q<1$ and equal to $\rho$ for larger $Q$ where $Q=k/(2k_f)$ \cite{hoye16} one finds the known answer Eq.~(II22)
\begin{equation}
F_{Lex}=-\frac{3\pi e^2 k_f}{2(2\pi)^3\varepsilon}_0\rho=12\rho\mu_f D I_{Lex},\quad \mbox{with} \quad I_{Lex}=-\frac{1}{4},
\label{52a}
\end{equation}
and where
\begin{eqnarray}
\nonumber
\rho&=&\frac{4\pi g}{3(2\pi)^3}k_f^3, \quad \frac{4\pi}{3}(r_s a_0)^3=\frac{1}{\rho}, \quad a_0=\frac{4\pi\varepsilon_0\hbar^2}{me^2},\\
 \mu_f&=&\frac{(\hbar k_f)^2}{2m}=\frac{50.1\,\mbox{eV}}{r_s^2},\quad \mbox{and} \quad D=\frac{m k_f}{2\pi^2\hbar^2}\frac{e^2}{\varepsilon_0(2k_f)^2}=0.082393\cdot r_s
\label{52b}
\end{eqnarray}
is used \cite{hoye16,lomba17}. Here $r_s$ is the dimensionless length parameter commonly used, $a_0$ is the Bohr radius, $k_f$ the Fermi wave vector, and $\mu_f$ is the Fermi energy which also is the chemical potential of the free electron gas at $T=0$. 

Now the exchange energy also may be found by rearranging integrations. This will be useful for its radiation exchange contribution by which analytical evaluation of it becomes possible too. At $T=0$ the Matsubara frequencies $K=2\pi n/\beta$ (with $n$ integer) become continuous, and we have ($2\pi\sum_n\rightarrow \beta \,dK$)
\begin{equation}
\tilde S(0,k)=\frac{1}{2\pi}\int\limits_{-\infty}^\infty\hat S(K,k)\,dK
\label{52c}
\end{equation}
where $\hat S(K,k)$ is given by integral (\ref{17}). First we write it as
\begin{equation}
\hat S(K,k)=\frac{2\pi}{(2\pi)^3}\frac{2m}{\hbar^2}\cdot 2\int\limits_{0<k^{'}<k_f}\int\limits_{-1}^1\frac{(k^2-2{\bf k}{\bf k'})}{G^2+(k^2-2{\bf k}{\bf k'})^2}\,d(-\cos\theta)\,k'\,^2\,dk'.
\label{36}
\end{equation}
Here the factor 2  is from the exchange of roles of $X$ and $Y$ in expression (\ref{14}) for $L$. Further $G=2mK/\hbar^2$ and ${\bf k}{\bf k'}=k k'\cos\theta$. The boundary $k_f$ for $k'$ is obvious for $k>2k_f$. For $k<2k_f$ it can still be used due to contributions that from symmetry must cancel where both $k'',k'<k_f$. Now quantities 
\begin{equation}
z=-\cos\theta,\quad t=\frac{k'}{k_f},\quad Q=\frac{k}{2k_f},\quad x=\frac{G}{2kk_f}=\frac{mK}{\hbar^2k k_f}=\frac{K}{4\mu_f Q}
\label{37}
\end{equation}
 are introduced to obtain
\begin{equation}
\hat S_L(K,k)=\hat S(K,k)=S_L(Q,x)=\frac{mk_f}{4\pi^2\hbar^2}I_L(Q,x),
\label{38}
\end{equation}
\begin{equation}
I_L(Q,x)=\frac{1}{Q}\int\limits_0^1\left[\int\limits_{-1}^1\frac{Q+tz}{x^2+(Q+tz)^2}\,dz\right]t^2\,dt.
\label{39}
\end{equation}
This together with the Coulomb interaction (\ref{19}) ($\sim 1/Q^2$) inserted in Eqs.~(\ref{52}) and (\ref{52a})   together with (\ref{52c}) gives ($\int_0^1\int_{-1}^1\, dz\, t^2\,dt=2/3$)
\begin{equation}
I_{Lex}=\frac{1}{\pi}\int\limits_0^\infty \left[\int\limits_{-\infty}^\infty I_L(Q,x)Q\,dx -\frac{2}{3}\right]\frac{1}{Q^2}Q^2\,dQ=\int\limits_0^1\int\limits_{-1}^1 \left[\int\limits_0^\infty I_L(Q)\,dQ\right] t^2\,dz\,dt,
\label{85}
\end{equation}
\begin{equation}
I_L(Q)=\frac{a}{|a|}-1, \quad a=Q+tz.
\label{87}
\end{equation}
Nonzero $I_L(Q)$ requires $Q+tz<0$, i.e. $Q<-tz$. Thus altogether (with $-tz>0$)
\begin{equation}
J_L=\int\limits_0^\infty I_L(Q)\,dQ=\int\limits_0^{-tz}(-2)\,dQ=2tz,
\label{88}
\end{equation}
\begin{equation}
I_{Lex}=\int\limits_0^1\int\limits_{-1}^0 J_L t^2\,dz\,dt=-\frac{1}{4}
\label{89}
\end{equation}
consistent with result (\ref{52a}).

\subsection{Radiation exchange energy}
\label{sec5}

The various elements of the correlation function matrix $P_{ij}$, given by (\ref{13}), differ by the factor
\begin{equation}
M_{ij}=(k_i^{''}-k_i^{'})(k_j^{''}-k_j^{'})=(k_i-2k_i^{'})(k_j-2k_j^{'})
\label{28}
\end{equation}
in their integrands. Here ${\bf k}^{''}={\bf k}-{\bf k}^{'}$ is used. With the $z$ axis directed along ${\bf k}=\{0,0,k\}$ there will be no contributions to integral (\ref{13}) for $i\neq j$. Further ($\sum_{ij}$)
\begin{equation}
k^2 M_{zz}=k_i k_j M_{ij}=(k^2-2{\bf k}{\bf k}^{'})^2=k^4-4k^2{\bf k}{\bf k}^{'}+4({\bf k}{\bf k}^{'})^{\,2},
\label{29}
\end{equation}
\begin{equation}
k^2(M_{xx}+M_{yy}+M_{zz})=k^2[(-2k'_x)^2+(-2k'_y)^2+(k-2k'_z)^2]=k^2(k^2-4{\bf k}{\bf k}^{'}+4k'^{\,2}),
\label{30}
\end{equation}
\begin{equation}
k^2 M_{xx}=k^2 M_{yy}=2(k^2k'^{\,2}-({\bf k}{\bf k}^{'})^{\,2}).
\label{31}
\end{equation}
Accordingly to compute the transverse correlation function $S_T=\hat S_T(K,k)$ the factor $k^2M_{zz}$ in the integral for the longitudinal one $S_L=\hat S_L(K,k)$ is replaced by $k^2 M_{xx}$. It occurs twice as it has a $y$ component too.

Thus with new quantities (\ref{37}) inserted in (\ref{29}) and (\ref{31}) we find the ratio
\begin{equation}
\frac{M_{xx}}{M_{zz}}=\frac{t^2(1-z^2)}{2(Q+tz)^2}.
\label{41}
\end{equation}
So relating this to the longitudinal (density-density) correlation function Eqs.~(\ref{38}) and (\ref{39}) the transverse correlation function becomes
\begin{equation}
S_T(Q,x)=\hat S_T(K,k)=\frac{m k_f}{4\pi^2 \hbar^2} I_T(Q,x),
\label{42}
\end{equation}
\begin{equation}
I_T(Q,x)=\frac{1}{2Q}\int\limits_0^1\left[\int\limits_{-1}^1\frac{t^2(1-z^2)}{[x^2+(Q+tz)^2](Q+tz)}\,dz\right]t^2\,dt.
\label{43}
\end{equation}

The exchange energy related to radiation can now be found by utilizing the steps for the electrostatic case where the Coulomb interaction is replaced by the radiating interaction from the $J_2$ term of interaction (\ref{18}). It is
\begin{equation}
\hat\psi_T (K,k)=\tilde\psi(k) h(K,k), \quad h(K,k)=\frac{-(\omega/c)^2}{k^2-(\omega/c)^2}.
\label{48}
\end{equation}
Here $\tilde\psi(k)$ is again the Coulomb interaction (\ref{18}). The frequency $\omega$ is related to the Matsubara frequency via (sign may depend upon convention)
\begin{equation}
K=i\hbar\omega.
\label{48a}
\end{equation}
With this and new quantities (\ref{37}) one finds
\begin{equation}
h(K,k)=\frac{K^2}{K^2+(c\hbar k)^2}=\frac{x^2}{x^2+Q_0^2}, \quad Q_0=\frac{mc}{\hbar k_f},
\label{50}
\end{equation}
where with relations (\ref{52b}) one has $Q_0^2=mc^2/(2\mu_f)$ and $\mu_f=50.1\,\rm{eV}/r_s^2$ by which 
\begin{equation}
Q_0=71.5\, r_s.
\label{51}
\end{equation}

Like Eq.~(\ref{52a}) the radiation excess energy per volume unit can be written as
\begin{equation}
F_{Tex}=2\cdot 12\rho\mu_f D I_{Tex}
\label{51a}
\end{equation}
where the factor 2 in front is due to equal contributions from both $x$ and $y$ directions.
With the modified correlation function and interaction Eq.~(\ref{85}) is modified into
\begin{equation}
I_{Tex}=\int\limits_0^\infty \left[\frac{1}{\pi}\int\limits_{-\infty}^\infty I_{T}(Q,x)\frac{1}{Q^2}\frac{x^2}{x^2+Q_0^2}Q\,dx\right]Q^2\,dQ=\int\limits_0^1\left[\frac{1}{2}\int\limits_{-1}^1 J(z,t)(1-z^2)\,dz\right]t^4\,dt
\label{70}
\end{equation}
\begin{equation}
J(z,t)=\int\limits_0^\infty \left[\frac{1}{\pi}\int\limits_{-\infty}^\infty\frac{x^2\,dx}{(x^2+a^2)(x^2+Q_0^2)a}\right]\,dQ, \quad a=Q+tz.
\label{71}
\end{equation}
Note that in expression (\ref{85}) a diverging self-energy part, proportional to density $\rho$, was subtracted. Due to its proportionality to $\rho$ it does not influence the equation of state, i.e. the pressure. This is not so for the present case where such divergence is not present. The result obtained is finite for $Q_0>0$.
By integration one finds
\begin{equation}
J(z,t)=\frac{1}{Q_0}\left[\ln(Q_0+|tz|)-\ln|tz|\right],
\label{74}
\end{equation}
\begin{equation}
I_{Tex}=\frac{1}{Q_0}\left[\frac{2}{15}\ln Q_0+\frac{46}{225}+2S\right]
\label{80}
\end{equation}
where with $\tau=1/Q_0$
\begin{eqnarray}
\nonumber
S&=&\frac{1}{15}\ln(1+\tau)+\frac{1}{8\tau}\left[\ln(1+\tau)-\tau\right]-\frac{1}{12\tau^3}\left[\ln(1+\tau)-\tau+\frac{1}{2}\tau^2-\frac{1}{3}\tau^3\right]\\
& &+\frac{1}{40\tau^5}\left[\ln(1+\tau)-\tau+\frac{1}{2}\tau^2-\frac{1}{3}\tau^3+\frac{1}{4}\tau^4-\frac{1}{5}\tau^5\right].
\label{83}
\end{eqnarray}
More details of the integrations can be found in the Appendix. For large $Q_0$, i.e. $\tau\rightarrow 0$ which means low density, the $S\sim\tau$ and can be neglected. For very small $Q_0$, i.e. $\tau\rightarrow\infty$, the two other terms in (\ref{83}) cancel and one is left with
\begin{equation}
I_{Tex}=-\frac{1}{4}\ln Q_0+\frac{1}{16}+\cdots.
\label{83a}
\end{equation}

One finds that the excess energy due to radiation has opposite sign compared to the electrostatic one. Their ratio is
\begin{equation}
2\frac{I_{Tex}}{I_{Lex}}=-8I_{Tex},
\label{83b}
\end{equation}
its factor 2 in accordance with Eq.~(\ref{51a}). The opposite sign of these excess contributions is as should expected. However, the radiation part exceeds the other for $Q_0$ below some value, which from Eq.~(\ref{51}) means small $r_s$, i.e. rather high densities.
The physical interpretation of this is not clear to us.

\section{Correlation energy}
\label{corr}
In this section we introduce the correlation energy due to  electrostatic and radiation
effects. 
\subsection{Electrostatic correlation energy}
\label{sec6}

The Coulomb interaction induces correlations. In the RPA  the resulting correlation function becomes \cite{hoye16,lomba17}
\begin{equation}
\hat\Gamma_L(K,k))=\frac{\hat S_L(K,k)}{1+\hat A_L(K)},
\label{90a}
\end{equation}
\begin{equation}
\hat A_L(K)=g\hat S_L(K,k)\tilde\psi(k)=D\frac{f_L(Q,x)}{Q^2}
\label{91a}
\end{equation}
with $D$ given by (\ref{52b}). The $f_L(Q,x)=I_L(Q,x)$ with the latter given by integral (\ref{39}). The known result is \cite{lein00,hoye16,lomba17}
\begin{eqnarray}
\nonumber
f_L(Q,x)=&-&\left[ \frac{Q^2-x^2-1}{4Q}\ln\left(\frac{x^2+(Q+1)^2}{x^2+(Q-1)^2}\right)\right.\\
&&\left.  -1+x\arctan\left(\frac{1+Q}{x}\right)+x\arctan\left(\frac{1-Q}{x}\right)\right].
\label{92a}
\end{eqnarray}
To obtain more accurate results an effective interaction was
introduced in Refs.~\cite{hoye16} and \cite{lomba17} ($Q=k/(2k_f)$
\begin{equation}
\tilde \psi_e(k)=\tilde\psi(k) L(Q)
\label{93a}
\end{equation}
where $L$ is a function
\begin{equation}
L(Q)=1-\frac{Q^2}{\kappa^2}+\cdots.
\label{94a}
\end{equation}
that cuts the interaction for small $r$ in $r$ space, i.e. $L\rightarrow 0$ as $Q\rightarrow \infty$. With effective interaction the $\hat A(K)$ in Eq.~(\ref{90a}) is replaced by
\begin{equation}
\hat A_e(K)=\hat A_L(K) L(Q).
\label{95a}
\end{equation}
The results depend somewhat upon the specific form of the function $L$ chosen. In Ref.~\cite{lomba17} the parameter $\kappa$ was optimized by thermodynamic self-consistency. One form, which was named Gaussian cut, gave very accurate results.

With the RPA the correlation energy per unit volume is given by \cite{hoye16,lomba17}
\begin{equation}
F_{Lcorr}=\frac{1}{2(2\pi)^32\pi}\int\int[\ln(1+\hat A_L(K)-\hat A_L(K)]\,dKd{\bf k}
\label{96a}
\end{equation}
 With the effective interaction in Eq.(\ref{91a}) and use of thermodynamic self-consistency the precise procedure to obtain $F_{Lcorr}$ is more involved. However, when the parameter $\kappa$ varies little, which was the case, the $\hat A_L(K)$ may with good accuracy be replaced by quantity (\ref{95a}). After that, the integrand of (\ref{96a}) is divided by $L(Q)$. By evaluations new variables of integration (\ref{37}) may be used along with relations (\ref{52b}) such that
\begin{equation}
\frac{dKd{\bf k}}{2(2\pi)^2 2\pi}=12 \rho\mu_f\frac{1}{\pi}\, dx\,Q^3\, dQ.
\label{98a}
\end{equation}

\subsection{Radiation correlation energy}
\label{sec7}

Induced correlations and free energy contribution from radiation can be found with precisely the same formalism as for the electrostatic part. With this the correlation function and interaction are replaced with the corresponding transverse parts. For the free energy with contributions from both $x$ and $y$ directions one also multiplies with a factor 2 consistent with Eq.~(\ref{27}) for $\langle J_2\rangle$.

Like Eqs.~(\ref{90a}) and (\ref{91a}) the transverse correlation function becomes
\begin{equation}
\hat\Gamma_T(K,k))=\frac{\hat S_T(K,k)}{1+\hat A_T(K)},
\label{100}
\end{equation}
\begin{equation}
\hat A_T(K)=g\hat S_T(K,k)\hat\psi_T(K,k)=D\frac{f_T(Q,x)}{Q^2}\frac{x^2}{x^2+Q_0^2},
\label{101}
\end{equation}
where $\hat S_T(K,k)$ is given by Eqs.~(\ref{42}) and (\ref{43}) and $\hat\psi_T(K,k)$ is given by Eqs.~(\ref{48}) and (\ref{50}). To obtain $f_T(Q,x)=I_T(Q,x)$ integral (\ref{43}) was performed by analytic evaluation on computer. We found
\begin{equation}
f_T(Q,x) = -\frac{3}{4}+ \frac{(1+x^2-Q^2 )}{2x}\left[\arctan\left(\frac{1+Q}{x}\right) +
\arctan\left(\frac{1-Q}{x}\right)\right]+\frac{\chi(Q,x)}{16Qx^2}
\label{102}
\end{equation}
with
\begin{eqnarray}
\nonumber
\chi(Q,x)&=& 2(1 - Q^2)^2\,\ln\left(\frac{1 + Q}{1 - Q}\right)\\
 &-& \left(-2Q^2 + Q^4 - 6Q^2x^2 + (1 + x^2)^2\right)\ln\left(\frac{(1
  + Q)^2 + x^2}{(1 - Q)^2 + x^2}\right).
\end{eqnarray}
Numerical integration of Eq.(\ref{102}) via Eq.(\ref{70}) produces
results for the transverse exchange energy consistent with  Eqs.~(\ref{80}) and (\ref{83}), as is
illustrated in Fig. \ref{Iex}. An additional benefit of Eq.~(\ref{102}) is that the wave vector distribution of the transverse exchange energy can be obtained. Then $I_T (Q,x)=f_T(Q,x)$ is inserted in Eq.~(\ref{70}) and the $x$ integration is performed. By that Eq.~(\ref{70}) turns into the form
\begin{equation}
I_{Tex}=\int\limits_0^\infty I_T(Q)\,dQ
\label{120}
\end{equation}
where $I_T(Q)$ is the result of its $x$ integration. Thus with Eq.~(\ref{51a}) the transverse exchange energy per particle and its wave vector distribution $\varepsilon_{Tex}(Q)$  become
\begin{equation}
f_{Tex}=\frac{F_{Tex}}{\rho}=\int\limits_0^\infty \varepsilon_{Tex}(Q)\,dQ, \quad \varepsilon_{Tex}(Q)
=24\mu_f DI_T(Q).
\label{121}
\end{equation}
In Fig.~\ref{epsq} the  $\varepsilon_{Tex}(Q)$ is compared with the corresponding distribution $\varepsilon_c(Q)$ for the longitudinal correlation energy from Ref.~\cite{lomba17}.

\begin{figure}[t]
  \includegraphics[width=12cm,clip]{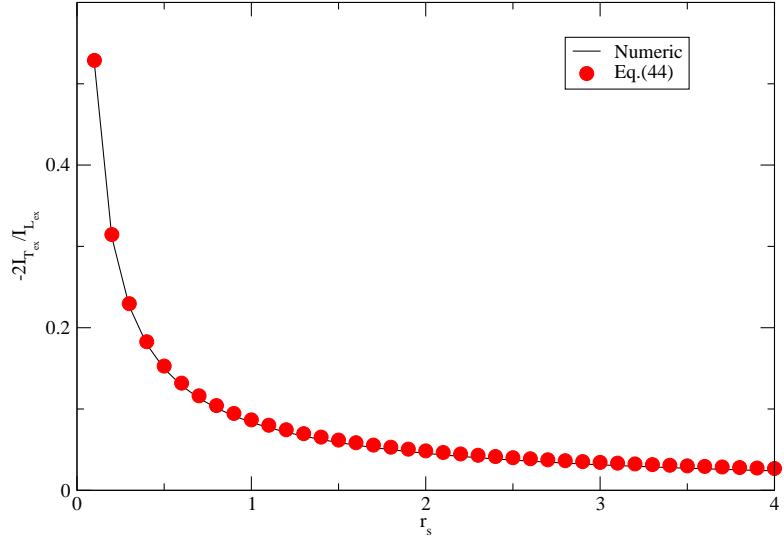}
  \caption{Relative magnitude of the radiation excess energy compared
    to the usual electrostatic contribution $-2I_{Tex}/I_{Lex}$ as computed numerically from
    Eq.(\ref{102}) and Eq.(\ref{70}) vs. the analytic expression of  Eqs.~(\ref{80}) and (\ref{83}). Note that $I_{Lex}$ is given by Eq.~(\ref{89}).}
   \label{Iex}  
\end{figure}

Likewise, from Eq.~(\ref{96a}) with the factor 2 included as in Eq.~(\ref{51a}), the transverse correlation energy per unit volume becomes
\begin{equation}
F_{Tcorr}=2\frac{1}{2(2\pi)^32\pi}\int\int[\ln(1+\hat A_T(K)-\hat A_T(K)]\,dKd{\bf k}.
\label{103}
\end{equation}
New variables of integration as given by Eq.~(\ref{98a}) are then used. The transverse correlation energy per particle follows from
\begin{equation}
f_{Tc}=\frac{F_{Tcorr}}{\rho}.
\label{105}
\end{equation}
From Fig.~\ref{ftc} it is seen that $f_{Tc}$ is very small.

The use of effective interaction with a cut function $L(Q)$ as
sketched below Eq.~(\ref{96a}) may also be used for
$F_{Tcorr}$. However, since the latter is expected to be a very small
correction to this already small quantity, it will be of little
significance anyway. So we have not investigated this further. 

\section{Dielectric constant}
\label{sec8}

The uniform electron gas can be thought as the free conduction
electrons of metals, and and so it can be regarded as a dielectric
fluid. In our derivations the electron gas has already been treated as
such a fluid. Now the dielectric properties are tightly connected to
the pair correlation function that can be seen as a resulting
interaction in the dielectric medium. 

The direct interaction between polarizations ${\bf s}_1$ and ${\bf s}_2$ in vacuum is given by expressions (\ref{18}) and (\ref{19}). In a dielectric medium the corresponding interaction should be
\begin{equation}
\hat \phi_\varepsilon(12,\omega)=\hat \psi_{\varepsilon L}(k,\omega) J_1+\hat \psi_{\varepsilon T}(k,\omega) J_2,
\label{110}
\end{equation}
\begin{equation}
\hat \psi_{\varepsilon L}
(k,\omega)=
\frac{\tilde\psi(k)}{\varepsilon_L}, \quad \hat \psi_{\varepsilon T}(k,\omega)=
\frac{\tilde\psi(k)}{\varepsilon_T}\,\frac{-\Omega^2}{k^2-\Omega^2}, \quad \Omega^2=\varepsilon_T\left(\frac{\omega}{c}\right)^2,
\label{111}
\end{equation}
where $\varepsilon_L$ and $\varepsilon_T$ are longitudinal and transverse dielectric constants. They are expected to be equal in the continuum limit of small wave vectors, $k\rightarrow 0$, and nonzero frequencies. For the electrostatic case $\omega=0$ the transverse part clearly does not contribute.

The interactions in a dielectric medium is given by the direct interaction (\ref{18}) plus induced contributions via the correlated particles of the dielectric medium. This involves the same convolutions used to obtain the resulting correlation functions (\ref{90a}) and (\ref{100}). Thus ($K=i\hbar\omega$)
\begin{equation}
\label{}
\end{equation}
\begin{equation}
\hat\psi_{\varepsilon L}(k,\omega)=\frac{\tilde\psi(k)}{1+\hat A_L(K)},\quad \hat\psi_{\varepsilon T}(k,\omega)=\frac{\hat\psi_T(K,k)}{1+\hat A_T(K)}.
\label{112}
\end{equation}
Comparing with Eq.~(\ref{111}) one finds
\begin{equation}
\varepsilon_L=1+\hat A_L(K)=1+g\hat S_L(K,k)\tilde\psi(k),\quad \varepsilon_T=1+g\hat S_T(K,k)\tilde\psi(k).
\label{113}
\end{equation}
The result for $\varepsilon_T$ follows by noting that it implies $1+\hat A_T(K)=1+(\varepsilon_T-1)h(K,k)$ with $h(K,k)$ given by (\ref{48}). One may also use the cut interaction (\ref{93a}) which implies multiplying the $\tilde\psi(k)$ in (\ref{113}) with $L(Q)$. Anyway, the concept of dielectric constant is mainly useful only in the long wavelength continuum limit, i.e. $k\rightarrow 0$, where the particle structure on the microscopic scale is not present. In this limit where $x\rightarrow \infty$ and $Q\rightarrow 0$ integral (\ref{39}) becomes
\begin{equation}
I_L(Q,x)=\frac{2}{3x^2}.
\label{45}
\end{equation}
When inserted in Eq.~(\ref{38}) and then inserted in  (\ref{113}) this results in
\begin{equation}
\varepsilon_L=1-\left(\frac{\omega_p}{\omega}\right)^2, \quad \omega_p^2=\frac{\rho e^2}{m \varepsilon_0},
\label{115}
\end{equation}
where Eqs.~(\ref{19}), (\ref{52b}), and (\ref{37}) are used. The $\omega_p$ is the plasma frequency. This result is the well known expression for the dielectric constant due to the free electrons in metals. 

The corresponding situation with integral (\ref{43}) for the transverse case is less straightforward. But one can write
\begin{equation}
\frac{t^2(1-z^2)}{Q+tz}=\frac{t^2}{Q+tz}+Q-tz-\frac{Q^2}{Q+tz}.
\label{116}
\end{equation}
By integration with respect to $z$ only the first and second terms contribute as the last one is of higher order in $Q$. The $z$-integration of these two terms gives $t\ln[(t+Q)/(t-Q))]+2Q\rightarrow 4Q$, $Q\rightarrow 0$. When used in integral (\ref{43}) the results is
\begin{equation}
I_T(Q,x)=\frac{2}{3x^2}.
\label{118}
\end{equation}
Accordingly in the limit $k\rightarrow 0$ ($\omega$ finite)
\begin{equation}
\varepsilon_T=\varepsilon_L.
\label{119}
\end{equation}

\section{Results}
\label{sec9}
The transverse correlation energy per particle (\ref{105}) can be evaluated by
numerical integration of Eq.~(\ref{103}), using the transformation
(\ref{98a}), and inserting Eqs.(\ref{101}) and (\ref{102}). The double
integration over Q and x is carried out using 10000 grid points for
each variable and grids $\Delta Q=\Delta x=0.01$. The numerical integration via Eq.~(\ref{102}) has
been checked against the analytical results for the evaluation of the
 radiation excess energy in Figure \ref{Iex}. 

 In Figure \ref{ftc} both the transverse excess energy due to
 radiation and the transverse correlation energy is presented and
 compared with the correlation 
    energy determined from the  Perdew-Wang
    fit \cite{PW} and the Gaussian cut approximation from
    Ref.~\cite{lomba17}. All these quantities are scaled with 
    the longitudinal excess energy (\ref{52a}). One can readily see
    that the transverse contribution  is mostly a small fraction of the
    longitudinal one. Relative to the latter it decays close to an inverse
    power of $r_s$ for decreasing density ($\rho\sim1/r_s^3$), but growing for very high densities
    according to Eq.~(\ref{83a}). As seen from the figure, the transverse correlation energy is very small compared to the excess part.
 \begin{figure}[t]
  \includegraphics[width=12cm,clip]{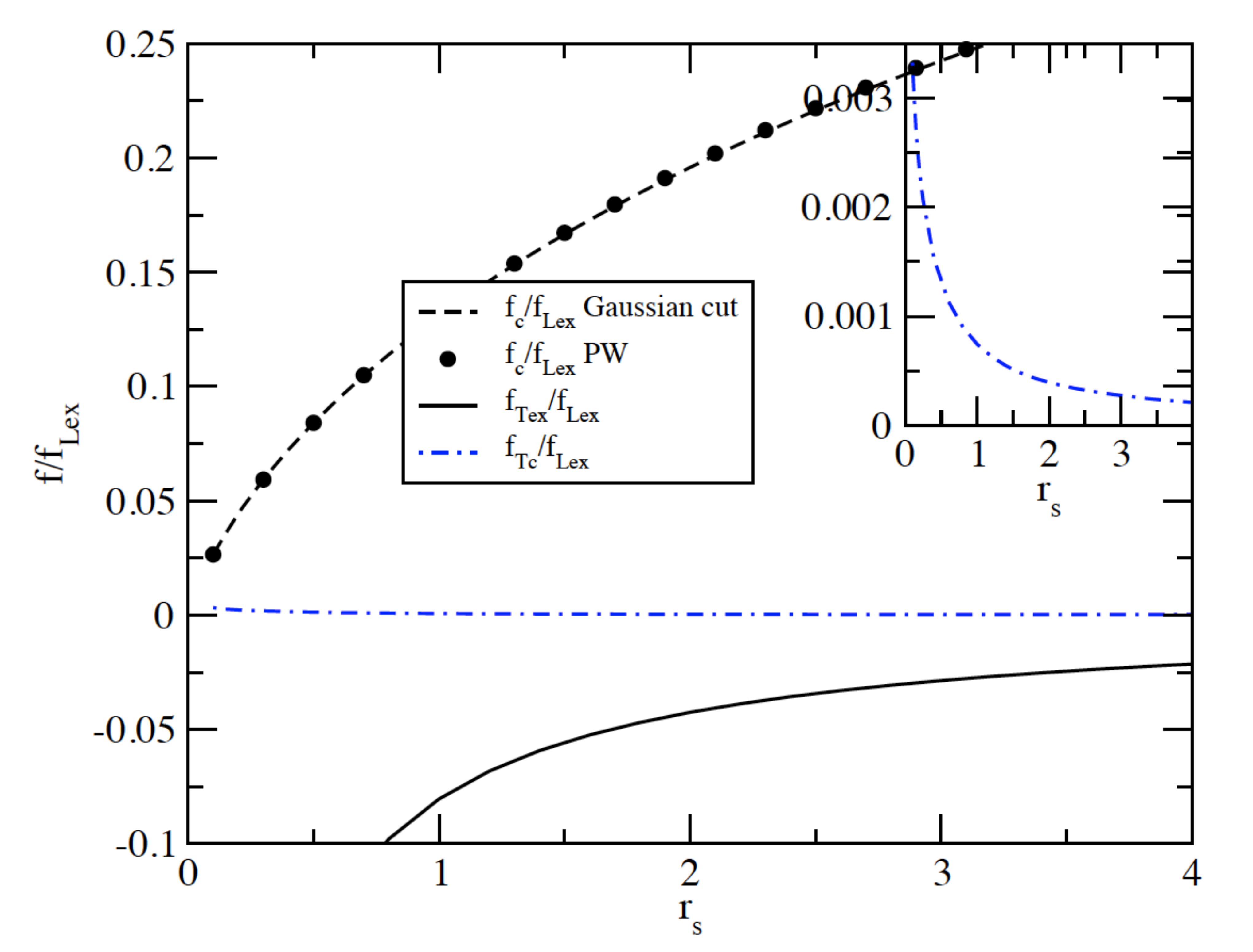}
  \caption{Transverse excess and correlation energies per particle $f_{Tex}=F_{Tex}/\rho$ and $f_{Tc}$ given by Eqs.~(\ref{51a}) and (\ref{105}) respectively. This is related to the Perdew-Wang
    fit \cite{PW} to the longitudinal electrostatic correlation energy per particle $f_c=F_{Lcorr}/\rho$ and the Gaussian cut approximation to it from
    Ref.~\cite{lomba17}. All curves for $f=f_c$, $f_{Tex}$, and $f_{Tc}$ respectively   are scaled with the
    longitudinal excess  energy  (\ref{52}) and (\ref{52a}) with $f_{Lex}=F_{Lex}/\rho$. The Gaussian cut $F_{Lcorr}$ was found as sketched  below Eq.~(\ref{96a}). \label{ftc} }
\end{figure}

The behavior of the transverse excess energy contribution from radiation is illustrated by the wave 
vector analysis of it in Fig.~\ref{epsq}. There the wave vector distribution $\varepsilon_{Tex}(Q)$
of Eq.~(\ref{121}) is plotted. It is compared with the corresponding analysis of the longitudinal electrostatic correlation energy from the 
Perdew-Wang
    fit \cite{PW} and the Gaussian cut approximation from
    Ref.~\cite{lomba17}. One can see that for $r_s$ characteristic of
    aluminum (one of the smallest, $\sim 2$) the correction due to radiation is hardly
    noticeable. However, for much higher densities ($r_s=0.5$) the
    correction is considerable where also values for $Q=k/(2k_f)>1$ give a significant contribution.
 \begin{figure}[t]
  \includegraphics[width=12cm,clip]{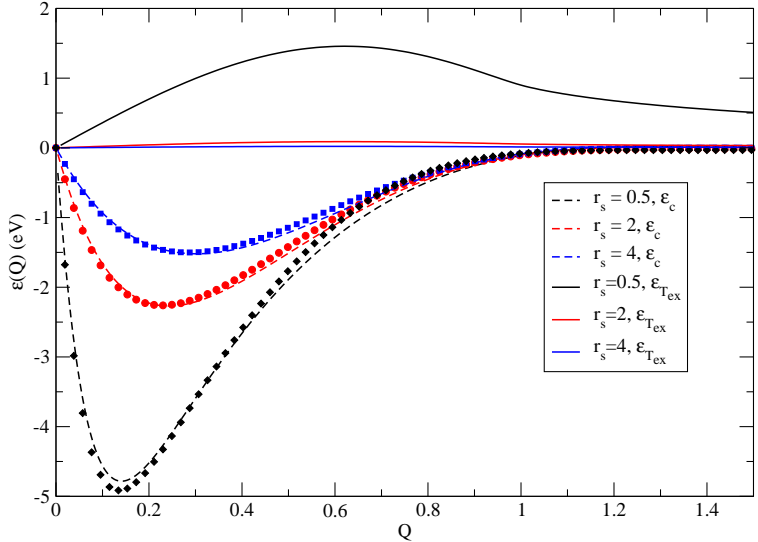}
  \caption{Wave vector analysis of the electron transverse excess energy due to radiation where its distribution $\varepsilon(Q)=\varepsilon_{Tex}(Q)$ from Eq.~(\ref{121}) is plotted (fully drawn curves).  This is compared with a corresponding plot of the longitudinal correlation energy $\varepsilon(Q)=\varepsilon_c(Q)$ computed
    from  the Perdew-Wang
    fit \cite{PW} (dots) and the Gaussian cut approximation from
    Ref.~\cite{lomba17} (dashed curves).\label{epsq} }
\end{figure}

\section*{Summary}
\label{secS}

In this work we have obtained radiation corrections to the (free) energy of the quantized uniform electron gas at $T=0$.
It is positive and adds to the negative electrostatic Coulomb energy to reduce it. In comparison its relative 
magnitude is small, except for rather large densities. It decreases rapidly with decreasing densities. 
The greater part of it is the transverse excess energy. This energy comes from the coupling of the radiating 
electromagnetic field to the current correlations of the free unperturbed fermion gas of electrons.
This coupling also induces transverse contributions to the correlation function that give the transverse correlation free energy.
But this free energy contribution turns out to be very small. Further the induced longitudinal electrostatic 
and transverse correlation functions
defines non-local longitudinal and transverse dielectric constants of the uniform electron gas.
As should be expected they are equal and in accordance with the well-known plasma relation  for oscillations
in the electron gas in the long wavelength limit ($k\rightarrow0$).

\section*{Appendix}
\label{secA}

Here details of the evaluation of integral (\ref{70}) is given. Expression (\ref{74}) for $J(z,t)$ makes the integrand symmetric around $z=0$ by which $z$ can be restricted to lie between 0 and 1. This is compensated by multiplication with 2 and the integral becomes
\begin{equation}
I_{Tex}=\int\limits_0^1\int\limits_0^1\frac{1}{Q_0}\left[\ln(Q_0+tz)-\ln(tz)\right](1-z^2)\,dz\,t^4\,dt.
\label{75}
\end{equation}
The last $\ln$ integral is easily obtained by partial integrations ($\int_0^1 u^n\ln u\, du=-1/(n+1)^2$)
\begin{equation}
-\int\limits_0^1\int\limits_0^1(\ln t+\ln z)(1-z^2)\,dz\,t^4\,dt=\frac{46}{225}. 
\label{76}
\end{equation}
The first $\ln$ integral will be more cumbersome by partial integrations. However, we choose to expand the $\ln$ term in the quantity $tz\tau$ where 
\begin{equation}
\tau=\frac{1}{Q_0}.
\label{77}
\end{equation}
\begin{equation}
\ln(Q_0+tz)=\ln Q_0+\sum\limits_{n=1}^\infty\frac{(tz\tau)^n}{n}(-1)^{n-1}.
\label{78}
\end{equation}
By  integration of the resulting power series the two $z$ integrations result into the factor
\begin{equation}
\frac{1}{n+1}-\frac{1}{n+3}=\frac{2}{(n+1)(n+3)},
\label{79}
\end{equation}
while the $t$  integration gives the factor $1/(n+5)$. Finally $\int_0^1\int_0^1(1-z^2)\,dz \,t^4\,dt=2/15$. Altogether we find 
\begin{equation}
I_{Tex}=\frac{1}{Q_0}\left(\frac{2}{15}\ln Q_0+\frac{46}{225}\right)+\frac{2}{Q_0}S,
\label{80b}
\end{equation}
\begin{equation}
S=\sum_{n=1}^\infty\frac{(-1)^{n-1}\tau^n}{n(n+1)(n+3)(n+5)}.
\label{81}
\end{equation}

The series for $S$ may be summed, and expansion in partial fraction gives
\begin{equation}
\frac{1}{n(n+1)(n+3)(n+5)}=\frac{1}{15n}-\frac{1}{8(n+1)}+\frac{1}{12(n+3)}-\frac{1}{40(n+5)}.
\label{82}
\end{equation}
With this four logarithmic series are obtained as given by expression (\ref{83}) for $S$.
\section*{Acknowledgment}
EL  acknowledges the support from the Agencia Estatal de
  Investigación and Fondo Europeo de Desarrollo Regional (FEDER) under
  grant No. FIS2017-89361-C3-2-P.
JSH acknowledges financial support from the Research Council of Norway, Project 250346.

\end{document}